\documentclass[prl,twocolumn]{revtex4}
\usepackage{graphicx}
\usepackage{amsmath}
\usepackage{amsfonts}
\usepackage{amssymb}
\begin{document}

\title{Large ferroelectric  polarization in\\
 the new  double perovskite NaLaMnWO$_{6}$\\
induced by non-polar instabilities}

\author{T. Fukushima\footnote{Present address: Graduate School of Engineering Science, Osaka University, 1-3 Machikaneyama, Toyonaka, Osaka 560-8531, Japan}}
\email{fuku@aquarius.mp.es.osaka-u.ac.jp}
\author{A. Stroppa}
\author{S. Picozzi}
\affiliation{%
Consiglio Nazionale delle Ricerche - Institute for Superconducting and 
Innovative Materials and Devices (CNR-SPIN), 67100 L'Aquila, Italy \\
}
\author{J. M. Perez-Mato}
\affiliation{Departmento de F\'{i}sica de la Materia Condensada, Facultad de 
Ciencia y Tecnolog\'{i}a, Universidad del Pa\'{i}s Vasco, Apartado 633, E-48080 
Bilbao, Spain
}%

\date{\today} 
              
\begin{abstract}
  
Based on density functional theory calculations and group theoretical analysis, 
we have studied   NaLaMnWO$_{6}$ compound  
which has been recently synthesized [Phys. Rev. B 79, 224428 (2009)] and 
belongs to the  $AA'BB'{\rm O}_{6}$ family of double perovskites.
At low temperature, the structure has monoclinic $P2_{1}$ symmetry, with layered 
ordering of the Na and La ions and rocksalt ordering of Mn and W ions. The Mn 
atoms show an antiferromagnetic (AFM)  collinear spin ordering, and the compound has 
been reported as a potential multiferroic. By comparing the low symmetry 
structure with a parent phase of  $P4/nmm$ symmetry, two distortion modes are 
found dominant.  They correspond to MnO$_{6}$ and WO$_{6}$ octahedron 
\textit{tilt} modes, often found in many simple perovskites. While in the latter 
these common tilting instabilities yield non-polar phases, in 
NaLaMnWO$_{6}$ the additional presence of the $A$-$A^{'}$ cation ordering is 
sufficient to make these rigid unit modes as a source of the 
ferroelectricity. Through a trilinear coupling with the two unstable tilting 
modes, a significant polar distortion is induced, although the system has  
no intrinsic polar instability. The calculated electric polarization resulting from 
this polar distortion is as large as $\sim$ 16 ${\mu}{\rm C/cm^{2}}$. Despite 
its secondary character, this polarization is coupled with the dominant tilting 
modes and its switching is bound to produce the switching of one of two tilts, 
enhancing in this way a possible interaction with the magnetic ordering. The 
transformation of common non-polar purely steric instabilities into sources of 
 ferroelectricity through a controlled modification of the parent 
structure, as done here by the cation ordering, is a phenomenon to be further 
explored.   
\end{abstract}

\pacs{.}%

\maketitle

Perovskite oxides are one of the most interesting and studied classes of 
inorganic compounds. They show many interesting properties such as 
ferroelectricity, ferromagnetism, superconductivity, colossal magnetoresistance, 
etc.\cite{TMO} The prototype structure is the simple $AB$O$_{3}$, where the 
three-dimensional framework of corner-sharing $B$O$_{6}$ octahedra form 12-
coordinate cavities occupied by the larger $A$ cations. The possibility of $A$- 
or $B$- substitution gives rise to a large compositional range with different 
properties and symmetries. A well studied case is the 50:50 substitution of $B$-
site in the double perovskites $A_{2}BB'{\rm O}_{6}$, where the $B$ and $B'$ 
cations order in a rock-salt fashion.\cite{Anderson,Lufaso} Analogously, 
ordering of the $A$-cations can also occur, showing a strong preference for a 
layered arrangement.\cite{Millange,Woodward,King2} Finally, 
\textit{simultaneous} $A$- and $B$-site cation ordering can also occur in 
compounds of type  $AA'BB'{\rm O}_{6}$.\cite{King1} These compounds are 
interesting from a structural point of view since they exhibit rocksalt ordering 
of $B$-site cations and layered ordering of $A$-site cations. However, they are  
relatively unexplored, because this type of ordering is very 
rare.\cite{King1,Knapp} Clearly, the presence of  four different distinct cation sites greatly increases the possibility of functional design in this class of compounds. 
NaLaMnWO$_{6}$ that has been recently synthesized\cite{King2} belongs to the 
 $A$ and  $B$ ordered double perovskites.
 Mn and W ions are octahedrally coordinated to O ions and these octahedra are strongly tilted: the tilting is represented as $a^{+}a^{+}c^{-}$ in terms of Glazer notation\cite{Glazer}.
It exhibits a polar crystal structure with monoclinic $P2_{1}$ symmetry and an 
AFM spin configuration with N\'{e}el temperature $T_{\rm N}=10$ K 
\cite{King1,King2,Knapp}. $P2_{1}$ is a \textit{polar} group where
  electric polarization is allowed. Therefore, the magnetic ordering at low 
temperature, suggests a possible multiferroic behavior \cite{King1,King2,Knapp}, 
showing both spontaneous ferroelectric \textit{and} magnetic 
order.\cite{Mostovoy,Ramesh,Picozzi,multiferroics}. 
Typical multiferroics belong to the group of  perovskite transition metal 
oxides, and include rare-earth manganites and ferrites (\textit{e.g.} 
TbMnO$_{3}$, HoMnO$_{3}$ HoMn$_{2}$O$_{5}$, LuFe$_{2}$O$_{4}$). Other examples 
are  bismuth compounds BiFeO$_{3}$ and BiMnO$_{3}$, non-oxides, such as 
BaNiF$_{4}$, and spinel chalcogenides, \textit{e.g.} 
ZnCr$_{2}$Se$_{4}$\cite{mf1,mf2,mf3,mf4,mf5,mf6,mf7,mf8,mf9,mf10,mf11,mf12,claude,mf13,mf14}.
 In these compounds ferroelectricity is either an independent phenomenon (like in 
BiFeO$_{3}$), where ferroelectricity can be caused by an intrinsic polar instability, or its existence is an induced effect of the magnetic ordering (as in 
TbMnO$_{3}$), where ferroelectricity is a secondary effect, i.e. {\it improper} ferroelectricity.

Very recently, new mechanisms for the coexistence and coupling of the two electric 
and magnetic ordering have been investigated. An intriguing possibility is that 
a coupling of non-polar instabilities can give rise to a polarization, as first theoretically predicted in Ref.\ \cite{trilinear}. Perez-Mato \textit{et al.}
studied SrBi$_{2}$Nb$_{2}$O$_{9}$ compound and found that
the ferroelectricity is due to the 
interplay of several degrees of freedom, surprisingly
not all of them associated to unstable or nearly-unstable modes. In particular, 
a coupling between polarization and two rotational modes has been described.\cite{perez-mato-PRB-2004-Aurivillius}
More recently, Bousquet \textit{et al.} has demonstrated that ferroelectricity is produced by local rotational modes in a SrTiO$_{3}$/PbTiO$_{3}$ superlattice.\cite{Bousquet} Following this work, Benedeck and Fennie
proposed that magnetoelectric coupling, weak ferromagnetism, and ferroelectricity can develop from the combination of two lattice rotations, neither of which produces ferroelectric properties individually.\cite{Fennie} In other words, separate lattice distortions can not only produce ferroelectricity, but also modify the magnetic order and favor magnetoelectricity.\cite{Fennie2,Fennie3,Fennie4}

 In this paper, we will demonstrate that, by means of {\it ab initio} density functional theory (DFT) calculations and group theoretical analysis, a similar mechanism  works  in the case of NaLaMnWO$_{6}$ producing a relatively large $\sim$ 16 ${\mu}{\rm C/cm^{2}}$ polarization:  two tilting instabilities are not enough to induce ferroelectric polarization by themself, but their coupling is sufficient to induce a polar instability and, therefore, ferroelectricity. To this end, we will show that the additional degree of freedom provided by the cation $A$-$A'$ ordering has an important role. 
The mechanism explained in the article by Fennie \textit{et al.}\cite{Fennie} and termed ``hybrid improper ferroelectricity'' is basically the same as the one discussed here, although for other type of systems. The calculated polarization is large here, due to the intrinsic softness of the polar mode in NaLaMnWO$_{6}$. The condensation of a hard mode through a trilinear coupling with two unstable modes is in fact a quite common mechanism in perovskite-related compounds and other materials \cite{Perez-Mato-ActaA-2010, perez-mato-PRB-2004-Aurivillius}, and can also be the origin of the simultaneous condensation of several distinct soft-modes, effectively acting as a single order parameter \cite{Etxebarria}

\begin{table}[b]
\begin{center}
\begin{tabular}{c@{\hspace{3mm}}c@{\hspace{2.5mm}}c@{\hspace{2.5mm}}c@{\hspace{3
mm}}ccc}
\hline \hline
Atom & & Exp.\cite{King2} & & & GGA+$U$ & \\
\hline
&$x$&$y$&$z$&$x$&$y$&$z$ \\
Na (2a) & 0.2514 & 0.2289 & -0.0008 & 0.2489 & 0.2223 & 0.0004 \\ 
La  (2a) & 0.2556 & 0.2755 & 0.4984 & 0.2574 & 0.2845 & 0.5003 \\
Mn (2a) & 0.7571 & 0.2507 & 0.2389 & 0.7517 & 0.2526 & 0.2392 \\
W (2a) & 0.7581 & 0.2494 & 0.7652 & 0.7547 & 0.2451 & 0.7664 \\
O$_{1}$ (2a) & 0.5345 & 0.5017 & 0.6944 & 0.5461 & 0.5118 & 0.6973 \\
O$_{2}$ (2a) & 0.5136 & 0.5364 & 0.3017 & 0.5222 & 0.5365 & 0.3021 \\
O$_{3}$ (2a) & -0.0205 & -0.0584 & 0.2250 & -0.0362 & -0.0681 & 0.2197 \\
O$_{4}$ (2a) & -0.0669 & -0.0358 & 0.7819 & -0.0692 & -0.0482 & 0.7840 \\
O$_{5}$ (2a) & 0.8279 & 0.2285 & 0.5111 & 0.8320 & 0.2252 & 0.5090 \\
O$_{6}$ (2a) & 0.6684 & 0.2565 & -0.0166 & 0.6664 & 0.2718 & -0.0127 \\
\hline \hline
\end{tabular}
\caption{Experimental\cite{King2} and theoretical optimized atomic coordinates 
of NaLaMnWO$_{6}$ with $P2_{1}$ crystal structure. The lattice constants are 
fixed to experimental values: $a=5.5717$ \AA, $b=5.5970$ \AA, $c=8.0155$ \AA, 
and $\beta=90.225^{\circ}$.}
\label{table1}
\end{center}
\end{table}

\begin{figure} [t]
\begin{center}
\includegraphics[width=8.5cm,clip]{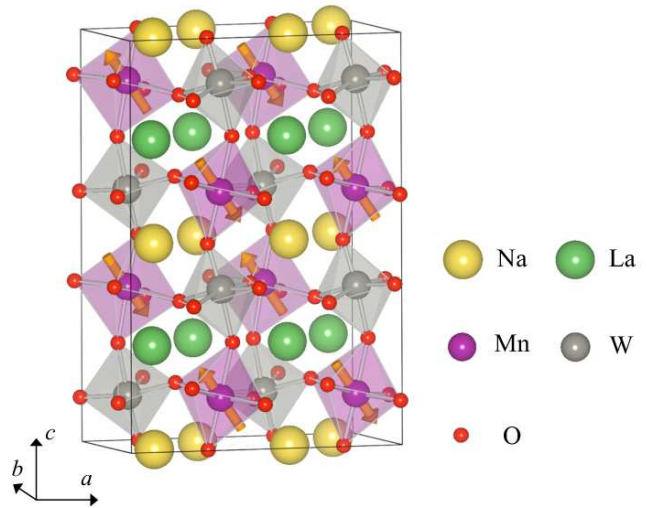}
\caption{Magnetic unit cell of NaLaMnWO$_{6}$ with  propagation vector 
$k=(0.5,0,0.5)$ which doubles the 
nuclear unit cell along $a$ and $c$ directions. The orange arrows indicate the 
spins of Mn atoms.}
\label{crystal}
\end{center}
\end{figure}

Electronic structure calculations and structural optimizations were performed by 
using  the ``Vienna {\it Ab initio} Simulation Package'' (VASP) and Projector 
Augmented Wave (PAW) pseudopotentials.\cite{Kresse} The Perdew-Becke-Erzenhof 
(PBE)  was employed for the exchange-correlation potential.\cite{Perdew} The  
DFT+$U$ method was used to properly take into account  correlation 
effects\cite{Dudarev}  using $U$=4 eV. We used   experimental lattice 
constants and we relaxed the internal coordinates\cite{King2}. The magnetic 
structure is an AFM collinear spin configuration with propagation 
vector $k=(0.5,0,0.5)$ (see Fig.\ \ref{crystal}). A $2{\times}4{\times}1$ 
Monkhost-Pack $k$-point grid in the Brillouin zone was used. The Berry phase 
approach \cite{Vanderbilt,Resta} was employed to calculate the electric 
polarization. 

\begin{figure} [t]
\begin{center}
\includegraphics[width=8cm,clip]{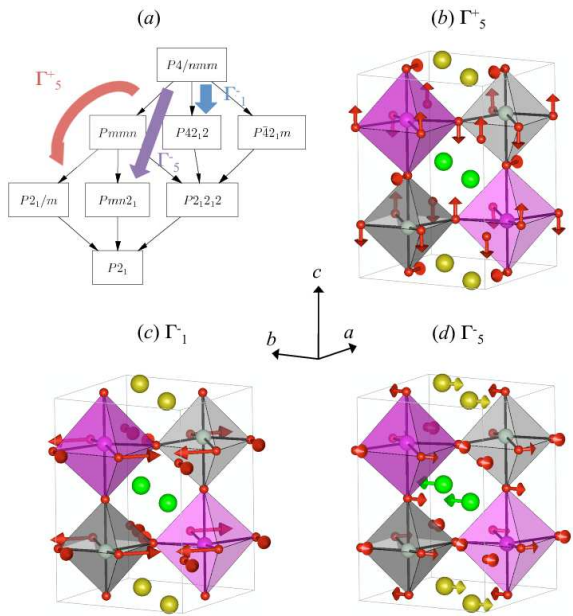}
\caption{(Color online) (a) Group-subgroup tree connecting the paraelectric 
$P4/nmm$ phase to ferroelectric $P2_{1}$ phase. (b), (c), and (d) are the 
pattern of  atomic displacements for the $\Gamma^{+}_{5}$, $\Gamma^{-}_{1}$, and 
$\Gamma^{-}_{5}$ distortion mode, respectively. Yellow, green, purple, grey, and 
red spheres indicate Na, La, Mn, W, and O atoms, respectively.}
\label{subgroup}
\end{center}
\end{figure}

A relaxed structure within the experimentally observed $P2_{1}$ symmetry was 
first calculated. The relaxed atomic positions are in good agreement with   
experimental values,\cite{King2} as shown in Table\ \ref{table1}. Mn-O and W-O 
corner sharing octahedra are strongly tilted and rotated due to the  
GdFeO$_{3}$-type distortion. It is well known that when the ionic radius of the 
$A$ and $A'$ cation is small ($r_{\rm Na}=1.22 {\rm \ \AA}$, $r_{\rm La}=1.24 
{\rm \ \AA}$\cite{Shannon}), this distortion is enhanced. In fact, in our case,  
the Goldschmidt tolerance factor, \textit{i.e.} (${\langle}r_{\rm 
Na,La}\rangle+r_{\rm O})/[\sqrt{2}({\langle}r_{\rm Mn,W}\rangle+r_{\rm 
O})]$=0.88, is smaller than 1. $\langle$Mn-O$\rangle$ and $\langle$W-O$\rangle$ 
bond lengths are 2.126 and 1.953 $\rm \AA$, respectively. The smaller 
$\langle$W-O$\rangle$ bond length is consistent with   experimental results, 
where W ions are strongly displaced from the centers of the octahedra due to a 
second order Jahn-Teller effect.\cite{King1}. 

The tiltings present in NaLaMnWO$_{6}$ can be described using the Glazer 
notation \cite{Glazer}, but it is preferable for our purposes to describe them 
within a comprehensive quantitative mode decomposition of the structure. 
\cite{Perez-Mato-ActaA-2010} The $P2_{1}$ structure of NaLaMnWO$_{6}$ can be 
considered as the distorted structure of a parent paraelectric $P4/nmm$ phase, 
which is the highest symmetry structure compatible with the mentioned order of 
the $A$ and $B$ sites. This idealized prototype structure can be found using the 
pseudosymmetry searching software \textit{PSEUDO}.\cite{Pseudo} 
Basically, the structure is  the ideal perovskite except for the ordering of the 
cations and there is no $B$O$_{6}$ and $B'$O$_{6}$ octahedral tilting or 
rotation. Using the symmetry mode analysis program {\it 
AMPLIMODES}\cite{Amplimodes}, one  can decompose the distortion relating the 
relaxed monoclinic structure with this prototypical
 structure into symmetry-adapted 
distortion modes. Three symmetry-breaking distortions can be distinguished with 
symmetries labelled as $\Gamma^{+}_{5}$, $\Gamma^{-}_{1}$ and $\Gamma^{-}_{5}$. Figure \ref{subgroup}(b, c and d) show  the pattern of atomic displacements 
associated with these three different distortion modes. The $\Gamma^{+}_{5}$ and 
$\Gamma^{-}_{1}$ distortion modes can be seen to be essentially two tilting 
modes of the octahedron framework.  They correspond to the 
  so-called rigid unit modes (RUMs)\cite{dove}, that are frequently 
unstable in simple perovskites. Their usual labels when described with respect 
to a $Pm\overline{3}m$ perovskite are $R4+$ and $M3+$ and would yield a 
distorted non-polar structure of $Pnma$ ($Pbnm$) symmetry. Here, the structure 
of the modes is the same, but the cation ordering implies that the  reference 
parent structure has a smaller tetragonal symmetry with a larger unit cell, so 
that they correspond to $\Gamma$ modes for this larger tetragonal supercell.
The third distortion mode labelled $\Gamma^{-}_{5}$ and shown in Figure 
\ref{subgroup}(d) is the polar mode frozen, which should be responsible for the 
ionic spontaneous polarization present in this compound. The atomic 
displacements for this mode are mostly along the crystallographic $b$ direction 
(polar axis). Na and La atoms displace in opposite directions by 0.055 and 
0.108\ \AA\ respectively, while   Mn and W atoms remain essentially fixed. 
The maximum oxygen displacement along the polar axis is 0.136\ \AA.

\begin{figure} [t]
\begin{center}
\includegraphics[width=8.5cm,clip]{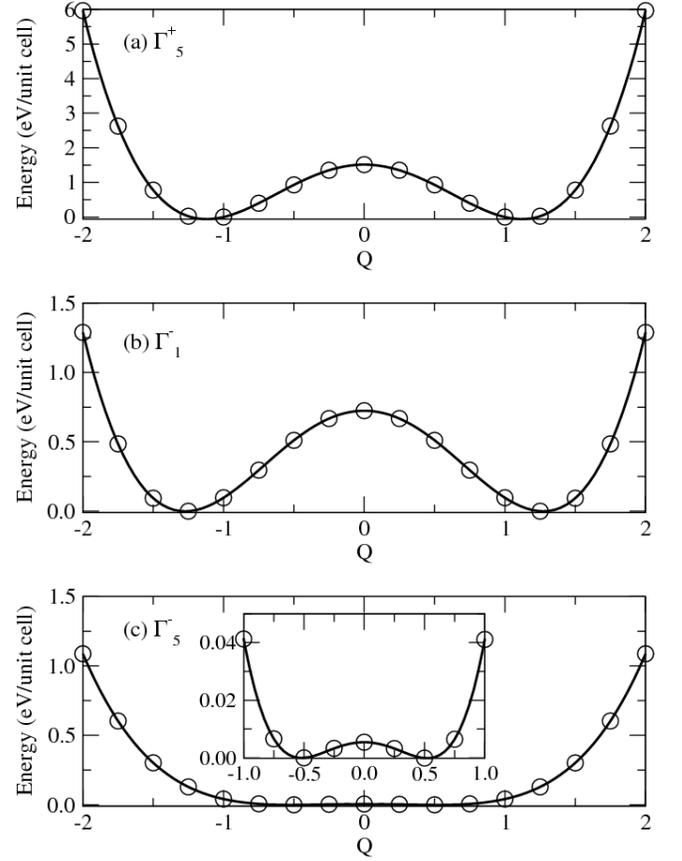}
\caption{(a), (b), and (c) are total energy variations as a function of the 
amplitudes of the $\Gamma^{+}_{5}$, $\Gamma^{-}_{1}$, and $\Gamma^{-}_{5}$ 
modes, respectively. The inset in (c) shows a zoom of $Q$ values in the [-1,+1] range.}
\label{phonon}
\end{center}
\end{figure}

The relative strength of each mode can be quantified by its amplitude, $Q$. 
\cite{Perez-Mato-ActaA-2010} In our case, $Q_{\Gamma^{+}_{5}}$ and Q$_{\Gamma^{-}_{1}}$  
are equal to 1.34 and 0.94 {\rm \AA}, respectively. On the other hand, the polar 
distortion $\Gamma^{-}_{5}$ has much smaller amplitude of 0.51 {\rm \AA}. 
The relative size of  symmetry-adapted mode amplitudes offer a valuable clue 
to interpret the mechanism giving rise to the polarization. The fact that 
$Q_{\Gamma^{+}_{5}}$ and Q$_{\Gamma^{-}_{1}}$ are much larger than $Q_{\Gamma^{-
}_{5}}$, suggests that the two first distortion modes are the \textit{primary} 
structural distortions with respect to the prototype phase, acting as order parameters, 
while the polar mode is a secondary induced distortion. Indeed, the 
$\Gamma^{+}_{5}$ and $\Gamma^{-}_{1}$ distortion modes lower the symmetry to 
$P2_{1}/m$ and $P42_{1}2$, respectively (\textit{isotropy} subgroups).
Note that none of them is a polar subgroup, although the second one breaks the 
inversion symmetry. This means that the structure resulting from these two 
dominant distortions, which is very close to the real structure, is a 
\textit{non-polar} structure. However, these two modes are sufficient to reduce 
the symmetry to $P2_{1}$, which allows the presence of a secondary induced 
\textit{polar} distortion. 
This can be understood considering the graph representing the 
group-subgroup relations connecting the high-symmetry $P4/nmm$ to the observed 
 $P2_{1}$, as shown in Fig.\ref{subgroup}(a). It can been seen that the 
intersection of the two mentioned isotropy subgroups is the observed $P2_{1}$ 
group, as it is their largest common subgroup. This means that the combination 
of these two tilting modes are able to induce a polar phase (although they don't
produce a polarization by themselves). It must be stressed that the original 
$R4+$ and $M3+$ unstable modes in pure $Pm\overline{3}m$ perovskites preserve in 
all cases a non-polar symmetry. It is only the fact that the cation ordering 
reduces the parent high symmetry to $P4/nmm$, which permits, in this compound, 
that two typical unstable non-polar perovskite tilting modes produce a polar 
structure, and, as a consequence, an improper ferroelectric.
To confirm this scenario we have investigated the stability of each of the three 
distortion modes present in the relaxed structure of NaLaMnWO$_{6}$. We have 
calculated the energy variation as a function of the amplitude of each mode, 
separately. The results are shown in Fig.\ \ref{phonon}. For the 
$\Gamma^{+}_{5}$ and $\Gamma^{-}_{1}$ modes, we have a characteristic double-
well shape, clearly indicating that they are unstable modes. 
On the other hand, the total energy variation of $\Gamma^{-}_{5}$ mode is a positive parabola, showing that the mode is stable. As shown in the inset of Fig.\ref{phonon}(c), there are minima around $Q=$0.5 and -0.5 for the $\Gamma^{-}_{5}$. In this sense, this mode is also unstable. However, the amplitude of the $\Gamma^{-}_{5}$ is 0.5086 $\rm \AA$ and corresponds to $Q=1$. Therefore, in order to achieve this amplitude,
 the $\Gamma^{-}_{5}$ distortion needs to combine with  primary distortions ($\Gamma^{+}_{5}$ and $\Gamma^{-}_{1}$ modes).   
   
The presence of the $\Gamma^{-}_{5}$ distortion with a quite significant 
amplitude can be explained as an induced effect through a symmetry-allowed 
anharmonic trilinear coupling with the primary tilting non-polar distortion modes
($\Gamma^{+}_{5}$ and $\Gamma^{-}_{1}$) $Q_{\Gamma^{-
}_{5}}Q_{\Gamma^{+}_{5}}Q_{\Gamma^{-}_{1}}$. This simple coupling is sufficient 
to explain the presence of a non-zero amplitude $Q_{\Gamma^{-
}_{5}}{\propto}(Q_{\Gamma^{+}_{5}}Q_{\Gamma^{-}_{1}})$ in order to minimize the 
energy, despite the mode itself being essentially stable. More importantly, this 
trilinear coupling will correlate the possible switching of polarization and 
tilting modes. For a fixed orientation of the polar axis, the number of domains 
is four and will be characterized by independent changes of sign of the two 
tilting modes, while the secondary polar mode correlates with them according to 
the proportionality law indicated above.  Therefore, 
polarization switching will imply necessarily a change of sign of one of the two 
rotational modes. This may be important to enhance the
 coupling of polarization with the magnetic ordering. 
Our results demonstrate that NaLaMnWO$_{6}$ can be considered an 
\textit{improper} ferroelectric.

\begin{figure} [t]
\begin{center}
\includegraphics[width=8.5cm,clip]{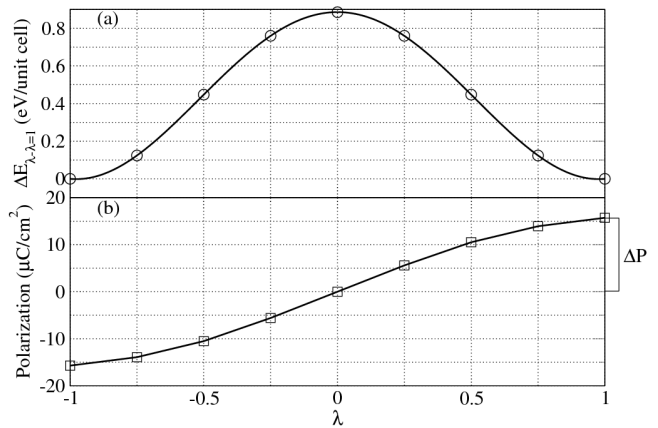}
\caption{(a) Total energy difference ${\Delta}E_{\lambda-\lambda_{=1}}$ as a 
function of parameter $\lambda$ (in eV/unit cell). (b) Adiabatic switching path 
along monoclinic $b$ direction between positive and negative ferroelectric 
states calculated by the Berry phase method (in $\mu{\rm C}/{\rm cm^{2}}$).}
\label{polarization}
\end{center}
\end{figure}

We now turn our attention to the estimate of the value of  spontaneous 
polarization in NaLaMnWO$_{6}$. For the calculation, the $P2_{1}/m$ phase 
resulting from the sole presence of the $\Gamma^{+}_{5}$ distortion is employed 
as a virtual paraelectric state and $\lambda$ is used as a scaling parameter of 
all ionic displacements, \textit{i.e.} $\lambda=0$ for the paraelectric state 
($P2_{1}/m$) and $\lambda=1$ for the  sum of the  two distortion components 
$\Gamma^{-}_{1}$+$\Gamma^{-}_{5}$ leading to the relaxed structure ($P2_{1}$). 
The negative $\mbox{\boldmath $P$}$ ($\lambda=-1$) state is built by ionic 
displacements opposite to the positive $\mbox{\boldmath $P$}$ state, with 
respect to the $\lambda=0$ structure. This corresponds to a correlated switch of 
the two distortion modes $\Gamma^{-}_{1}$ and $\Gamma^{-}_{5}$, as discussed 
above. Figure \ref{polarization} shows (a) the total energy difference and (b) 
the variation of  polarization as a function of $\lambda$. The total energy 
difference ${\Delta}E_{\lambda-\lambda_{=1}}=E_{\lambda}-E_{\lambda=1}$ shows 
convex and symmetrical shape between positive and negative $\mbox{\boldmath 
$P$}$ states, thus reproducing the bi-stable shape characteristic of  ferroelectric materials. The energy barrier between the paraelectric and ferroelectric states ${\Delta}E_{\lambda_{=0}-\lambda_{=1}}{\simeq}890$ meV/unit 
cell corresponding to 110 meV/formula unit (f.u.) is two times larger than in 
the YMnO$_{3}$ case, where ${\Delta}E{\simeq}60$ meV/f.u.\cite{Aken} Spontaneous 
ferroelectric polarization ${\Delta}\mbox{\boldmath $P$}$ calculated by the 
Berry phase method is $\sim$ 16 ${\mu}{\rm C/cm^{2}}$, as shown in 
Fig.\ref{polarization}(b). This value is  very close to the polarization value $\sim$ 14 ${\mu}{\rm C/cm^{2}}$ 
obtained by the point charge model with nominal charges (i.e. +1$e$ for Na, 
+3$e$ for La, +2$e$ for Mn, +6$e$ for W, and -2$e$ for O). This means that 
electronic effects are are negligible and  ionic displacements are the main 
contribution to  ferroelectric polarization. The relatively large value of polarization is consistent with the softness of the polar mode shown in 
Fig.\ \ref{phonon}. We have also checked that, by reducing the on-site Coulomb 
interaction $U$ to zero in our calculation, the polarization is not 
significantly affected.  This suggests that  Mn-3$d$ states do not play an 
important role in the ferroelectricity in NaLaMnWO$_{6}$.

Finally, we feel important to discuss two points, although they 
 go beyond the purpose of the present study. However, we believe that they  may stimulate further studies in this interesting class of compounds.
First, we have argued that the primary order parameters driving the transition in this compound are the two rotation modes, and not the polar mode, which happens to also be unstable.  For example, as in the Bousquet's work on PbTiO3/SrTiO3,\cite{Bousquet}
one rotation mode completely suppresses the other rotation mode, and it is required that the polar mode and a rotation mode be present in order for the second rotation mode to freeze in, \textit{i.e.}, the two rotation modes actually compete with each other and only one wins. In the work of Benedek and Fennie on Ca3Mn2O7,\cite{Fennie2} the polar mode is hard and the two rotation modes are compatible with each other, \textit{i.e.}, true rotation driven ferroelectricity. In our case, we did not investigate  whether the two rotation modes compete with each other or they are compatible: to properly address 
this question, a detailed  study of the anharmonic couplings is needed.
Second, a peculiar property
arise by studying the magnetoelectric coupling in this compound. 
By symmetry considerations, not reported here, one can show that:
i) a weak AFM magnetic moment M$_{y}$ should exist
 in this compound with the same propagation vector of the M$_{x}$ and M$_{z}$ components;
ii) the magnitude of the M$_{y}$ component should be  proportional to the
product of the amplitudes of the polar distortion and of the primary
frozen magnetic mode in the plane $xz$. This means that one should be able to see that this $y$ antiferromagnetic 
component switches by  switching the  sign of the polar mode and one  could expect
that it increases linearly with an increase of the amplitude of
the polar mode. In this way, one can simulate the effect of an electric
field along $y$, and if the $y$-AF component
responds linearly, this would correspond to a quite peculiar
\textit{magnetoelectric effect}: namely an AF ordering responding linearly to an
electric field. These considerations 
reinforce the idea of interesting physics in this novel class of compounds.

In conclusion, we have studied a new multiferroic material, NaLaMnWO$_{6}$, and 
shown that the compound is an \textit{improper} ferroelectric. The estimated 
ferroelectric polarization is $\sim$ 16 ${\mu}{\rm C/cm^{2}}$, a relatively high 
value compared with ferroelectrics of this type. A negligible  polar instability 
does  exist [see inset of Fig.\ \ref{phonon} (c)], but the additional A cation layer ordering of this double 
perovskite is sufficient to make ferroelectrically active some tilting modes of 
the octahedra that in simple perovskites and in B-ordered double perovskites 
only give place to non-polar phases. Through a trilinear coupling with the two 
unstable tilting modes, a significant polarization is induced. We hope that this 
study will stimulate further investigation of cation ordering  as a tool to convert ubiquitous well-known steric non-polar 
instabilities into mechanisms for producing improper ferroelectrics, as well as new 
multiferroics.

\acknowledgments
The research leading to these results has received funding from the European 
Research Council under the European Community's 7th Framework Programme 
(FP7/2007-2013) / ERC grant agreement n. 203523. Computational support by CASPUR 
supercomputing center (Rome) is acknowledged. The crystal structures in this 
paper are plotted by using the software {\it FP{\_}Studio} included in the {\it 
FullProf} suite.\cite{Fullprof1} A.S. greatly acknowledges the visiting period at 
University of the Basque country in Bilbao, where part of this work was done
and the warm hospitality of the members of this Institute.  Finally, 
we would like to thank very much  the anonymous referee for his/her
 useful comments.

\bibliography{biblio.bib}
\end{document}